\documentclass[pra,twocolumn,floatfix,superscriptaddress,aps]{revtex4}
\usepackage{graphicx}
\usepackage[usenames]{color}

\usepackage{amssymb}
\begin{document}

\title{A   self-bound   matter-wave boson-fermion quantum ball}

\author{ S. K. Adhikari}
 
\address{
Instituto de F\'{\i}sica Te\'orica, UNESP - Universidade Estadual Paulista, 01.140-070 S\~ao Paulo, S\~ao Paulo, Brazil
} 
%
\vspace{10pt}
%

\begin{abstract}

We  demonstrate the possibility of creating a self-bound   stable  three-dimensional matter-wave spherical  boson-fermion 
quantum ball  in the presence of an attractive boson-fermion interaction
and a
 small
repulsive three-boson interaction.  The three-boson interaction could 
be attractive or repulsive whereas the fermions are taken to be in a fully-paired super-fluid state in the Bardeen-Cooper-Schreifer ({ quasi-noninteracting weak-coupling})  limit.  
 We also include the Lee-Huang-Yang (LHY) correction  to a repulsive bosonic interaction term.     
The repulsive three-boson interaction and the LHY correction can stop a global  collapse while acting 
jointly or separately.  
The present study is based on   a  mean-field   model,  where the bosons are subject to a Gross-Pitaevskii (GP) Lagrangian functional and 
the fully-paired fermions are described by a Galilean-invariant density functional Lagrangian.
The boson-fermion interaction is taken to be the mean-field Hartree interaction, quite similar to the interaction term in the  GP equation.   
The study is illustrated by a variational and a numerical solution of the mean-field model for 
the boson-fermion $^7$Li-$^6$Li system.

\noindent{\it Keywords:\/} Bose-Einstein condensate, Superfluid fermion, soliton

\end{abstract}
\pacs{03.75.Mn, 03.75.Ss, 03.75Hh}


\maketitle

\section{Introduction}

A self-bound matter-wave  bright soliton  can travel with  a constant
velocity in one-dimension (1D) \cite{sol}, while maintaining its shape, due to a balance between 
 defocusing forces and 
nonlinear attraction. Solitons have been observed in diverse systems obeying classical and quantum dynamics, such as, in  water wave, nonlinear optics \cite{book} and Bose-Einstein condensate (BEC) \cite{becsol} among 
others. The 1D soliton could be analytic with energy and momentum conservation necessary to maintain its shape during propagation. However, such a soliton cannot be realized in three dimensions (3D) {  in the mean-field weak-coupling  Gross-Pitaevskii (GP) limit} 
 due to a collapse instability for attractive interaction \cite{sol,book}.

On the theoretical front Petrov  \cite{petrov} demonstrated the possibility of a 3D binary BEC droplet in the presence of an inter-species attraction and an intra-species repulsion with a Lee-Huang-Yang (LHY) correction \cite{LHY}. The possibility of forming a binary 1D BEC soliton with intra-species repulsion and inter-species attraction 
was suggested before \cite{PA}.
 In the presence of a
repulsive three-body interaction  the  statics and dynamics of a BEC quantum ball 
were studied in details recently
\cite{adhikari}
employing the numerical and variational solutions of a mean-field model. A droplet can also be realized in a  spin-orbit-  \cite{sandeep} or Rabi-coupled \cite{rabi}
multi-component spinor BEC.
 On the experimental front, a BEC droplet has been observed \cite{pfau} in a  dipolar dysprosium BEC with a repulsive short-range contact interaction.  Later, the formation of the dipolar droplet has been explained \cite{blakie}  by  a LHY correction to the short-range contact interaction.  More recently, a binary BEC droplet 
has been observed in the presence of a repulsive intra-species interaction and an attractive inter-species interaction  \cite{leticia,italy} and its formation  was explained by including a LHY-type correction term to the intra-species repulsion.

We demonstrate that it is possible to bind a large number of spin-1/2 fermions in a self-bound 3D boson-fermion  super-fluid  quantum ball at zero temperature   in the presence of
an attractive boson-fermion interaction and a repulsive   three-boson interaction together with the LHY correction for a repulsive boson-boson interaction. 
 We prefer the name quantum ball over droplet 
for the localized boson-fermion state after establishing the robustness of such a bosonic state 
to maintain the spherical ball-like structure 
after collision \cite{adhikari}, in contrast to easily deformable liquid droplets. 
Due to Pauli repulsion it is difficult to bind the fermions: the bosons with an attractive inter-species interaction 
act like a glue to bind the fermions.  The possibility  of binding fermions in  a 1D boson-fermion mixture without a trap
in  the presence of inter-species attraction 
was suggested theoretically \cite{1dbfth},  and later  realized experimentally \cite{1dbfex}.   In this study, we take the fermions  to be fully paired in a 
{  quasi-noninteracting}
weak-coupling super-fluid Bardeen-Cooper-Schrieffer (BCS) state, although this condition is not required for binding; all fermions in a spin-polarized state can also be bound in a boson-fermion quantum ball. 
The repulsive three-boson interaction and its LHY correction lead to terms with a  higher order nonlinearity in the dynamical ``mean-field" boson-fermion equation, compared to the nonlinearity resulting from the  boson-boson interaction,  and create a strong repulsive core at the origin and hence stop a global collapse of the boson-fermion mixture and stabilize  the quantum ball. 

We consider a numerical and a variational solution of a mean-field model for the formation of the boson-fermion quantum ball.   The Lagrangian functional of the bosons is taken as in the Gross-Pitaevskii 
(GP) Lagrangian functional including a three-boson interaction term and a LHY correction for a repulsive  boson-boson interaction and that of the fermions is taken as a Galilean invariant density functional Lagrangian \cite{bfmix}. The boson-fermion interaction is taken as the interaction term in the GP Lagrangian functional \cite{mix}. The Euler-Lagrange equations for the Lagrangian functional lead to a coupled set of equations employed in this study.  We illustrate the formation of a boson-fermion quantum ball  in the $^7$Li-$^6$Li mixture using realistic values of different parameters.

In Sec. II  the mean-field model for the boson-fermion mixture is developed.
A time-dependent, analytic,  Euler-Lagrange Gaussian variational approximation 
of the model is also presented. 
The results of numerical calculation are shown in Sec. III.  
Finally, in Sec. IV we present a brief summary of our findings.

\section{Analytic model for a boson-fermion quantum ball}

We consider a binary boson-fermion super-fluid  mixture  at zero temperature   interacting via  inter- and intra-species interactions with the 
mass and number of the two species $ i=1,2,$
denoted by $m_i, N_i,$ respectively. The first species ($^{7}$Li) 
is taken to be bosons 
 while the second species ($^{6}$Li) 
fermions.  The spin-half fermions are assumed to be fully paired with an equal number of 
spin-up and -down atoms.
 We start by writing the   Lagrangian density of the system  
\begin{eqnarray}{\cal L}&=&
\Big[
\sum_i  {\mbox i}\hbar \frac{N_i}{2}(\phi_i\dot \phi_i^*- \phi_i^* \dot \phi_i)
+ \frac{N_1\hbar^2}{2m_1}|\nabla \phi_1|^2 
\nonumber \\
& +&\frac{N_2\hbar^2}{8m_2} |\nabla \phi_2|^2+\frac{1}{3}\frac{\hbar N_1^3 K_3}{2} |\phi_1|^6
+\frac{1}{2} \frac{4\pi\hbar^2 a_1}{m_1} N_1^2|\phi_1|^4
\nonumber \\ &+&\frac{2}{5} \frac{2\hbar^2}{m_1}\pi   \alpha a_1^{5/2} N_1^{5/2}    |\phi_1|^5
+\frac{1}{2}
4\pi a_{12}N_1 N_2 \frac{\hbar^2}{m_R}|\phi_1|^2   |\phi_2|^2
\nonumber \\ &
+&\frac{3}{5} \frac{\hbar^2  }{2m_2}N_2(3\pi^2N_2)^{2/3}|\phi_2|^{10/3}  \Big], \quad  {\mbox i}=\sqrt{-1},
\label{lagden}
\end{eqnarray}
where $a_1$ is the scattering length of bosons (component 1), $a_{12}$ is the boson-fermion scattering length,  $m_R=m_1m_2/(m_1+m_2)$ is the boson-fermion reduced mass  and  the overhead dot denotes time derivative. In  (\ref{lagden}) the first term on the right is the usual time-dependent term \cite{bfmix,mfb2}, the second and the third terms represent the kinetic energies of bosons and fermions, respectively \cite{bfmix},  the term containing $K_3$ is the three-boson interaction term.    The prefactor $N_2\hbar^2/8m_2$ in the fermion kinetic energy guarantees 
Galilean invariance of the Lagrangian \cite{bfmix}.
The next term proportional to $a_1$ is the interaction energy of bosons and that proportional to $a_{12}$ is the boson-fermion interaction energy. 
The term containing $\alpha \equiv 64/(3\sqrt \pi)$    represents the beyond-mean-field  LHY correction to the repulsive bosonic intra-atomic interaction ($a_1>0$).   
 The fermions are assumed to be { quasi-noninteracting} in a completely full Fermi sea 
  and contributes the term proportional to $|\phi_2|^{10/3}$ in  (\ref{lagden}), which is just 
the static kinetic energy of all the fermions \cite{bfmix}.  
 Both the three-body and the LHY  terms have higher-order nonlinearity  compared to the two-body interaction term, viz. the term  containing $a_1$ in  (\ref{lagden}).  These terms  with a positive  real part of  $K_3$ 
guarantee a large positive energy near the origin ${\bf r}=0$ and 
stop the collapse of the system.

It is convenient to write a dimensionless form of expression (\ref{lagden})  as 
\begin{eqnarray}{\cal L}&=&
\Big[
\sum_i  {\mbox i}\frac{N_i}{2}(\phi_i\dot \phi_i^*- \phi_i^* \dot \phi_i)
+ \frac{N_1}{2}|\nabla \phi_1|^2+\frac{m_{1}}{8m_2} N_2|\nabla \phi_2|^2
\nonumber \\
& 
+& 2\pi a_1 N_1^2|\phi_1|^4
+ \frac{4}{5}\pi   \alpha a_1^{5/2} N_1^{5/2}    |\phi_1|^5+\frac{N_1^3 K_3}{6} |\phi_1|^6
\nonumber \\ &
+ &\frac{3m_1}{10m_2}N_2(3\pi^2N_2)^{2/3}|\phi_2|^{10/3}  
\nonumber \\ &
+ & 2\pi a_{12}N_1 N_2 \frac{m_1}{m_R}|\phi_1|^2   |\phi_2|^2\Big], \quad  {\mbox i}=\sqrt{-1},
\label{lag}
\end{eqnarray}
where  length is expressed in units of  a fixed length 
 $l  $,    density 
$|\phi_i|^2$ in units of $l^{-3}$,  time in units of $ 
t_0=m_1 l^2/\hbar$, energy in units of
 $\hbar^2/m_1 l^2$
and $K_3$ in units of $\hbar l^4/m_1$. The wave functions are normalized as $\int |\phi_i|^2 d{\bf r} = 1.$

 With  Lagrangian density  (\ref{lag}) the dynamics  for the binary boson-fermion mixture  is governed by the Euler-Lagrange equations 
\begin{equation}\label{EL}
\frac{d}{dt}\frac{\partial {\cal L}}{\partial \dot \psi_i^*}=  \frac{\partial {\cal L}}{\partial  \psi_i^*} .
\end{equation}
In explicit notation  (\ref{EL})  become  \cite{mfb2}
\begin{eqnarray}&  &\,
{\mbox i} \frac{\partial \phi_1({\bf r},t)}{\partial t}=
{\Big [}  -\frac{\nabla^2}{2 }
+ 4\pi a_1 N_1 \vert \phi_1 \vert^2 +\frac{K_3 N_1^2}{2} |\phi_1|^4
 \nonumber\\ &  &\, %
+  2\pi \alpha a_1^{5/2} N_1^{3/2}  |\phi_1|^3
+ \frac{2\pi m_1 a_{12} N_2}{m_R}  \vert \phi_2 \vert^2
{\Big ]}  \phi_1({\bf r},t),
\label{eq3}\\
& &\,
{\mbox i} \frac{\partial \phi_2({\bf r},t)}{\partial t}={\Big [}  
- \frac{m_1 \nabla^2}{8m_2} +\frac{m_1}{2m_2}  (3\pi^2 N_2)^{2/3} |\phi_2| ^{4/3} 
 \nonumber \\ & &\,
+ \frac{2\pi m_1 a_{12} N_1}{m_R}  \vert \phi_1 \vert^2
{\Big ]}  \phi_2({\bf r},t).
\label{eq4}
\end{eqnarray}

Convenient analytic variational approximation to  (\ref{eq3}) and (\ref{eq4}) can be obtained with the following Gaussian 
ansatz for the wave functions 
\cite{pg,mfb2}
\begin{eqnarray}\label{anz}
\phi_i({\bf r},t)=\frac{\pi^{-3/4}}{w_{ i}(t)\sqrt{w_{ i}(t)}}\exp\Big[-\frac{r^2}
{2w_{ i}^2(t)}+\mathrm{i}\beta_i(t)r^2\Big]
\end{eqnarray}
where $w_{ i}$ are the widths and $\beta_i$ 
 are additional 
variational parameters, called chirps. The effective Lagrangian for the binary system $ L=\int d{\bf r}{\cal L }$ is 
\begin{eqnarray}
 L
&= &
\sum_{i=1}^2\frac{N_i}{2}3w_{ i}^2\dot \beta_i  +\frac{N_1}{2}
\biggr[\frac{3}{2w_{ 1}^2} +6w_{ 1}^2\beta_1^2 \biggr]
\nonumber \\
&  +&\frac{N_2m_{1}}{8m_2}\biggr[\frac{3}{2w_{ 2}^2}
+ 6w_{ 2}^2\beta_2^2
 \biggr]+\frac{N_1^2a_1}
{\sqrt{2\pi}w_{ 1}^3} 
\nonumber\\ 
& +&\frac{8\sqrt 2}{25\sqrt 5} \frac{\alpha a_1^{5/2} N_1^{5/2}}{\pi^{5/4}w_1^{9/2}}
+\frac{N_1^3K_3}{18\sqrt{3}\pi^3 w_1^6}
\nonumber\\
 & 
+&\frac{9\sqrt 3 m_{1}(3 \pi^2 N_2)^{2/3} N_2}{50\sqrt 5  m_2\pi w_2^2} 
+\frac{2a_{12}m_1N_1N_2}{\sqrt \pi m_R
 (w_1^2+w_2^2)^{3/2}}    .      
\end{eqnarray}  
The repulsive three-boson $K_3$-dependent  term with a $1/w_1^6$ divergence and the LHY two-boson $\alpha$-dependent term with a
 $1/w_1^{9/2}$ divergence at the origin ($w_1=w_2=0$)  create a repulsive core in the Lagrangian $L(w_1,w_2)$  which stops the global collapse.

The four Euler-Lagrange variational equations of the effective Lagrangian $L$ 
for the four variational parameters $\alpha \equiv w_1,w_2, \beta_1, \beta_2$
\begin{equation}
\frac{d}{dt}\frac{\partial }{\partial \dot \alpha}= \frac{\partial L}{\partial \alpha},
\end{equation}
  can be simplified to yield 
the following coupled ordinary differential equations for the widths, $w_i$
 in usual fashion \cite{pg}
\begin{eqnarray}\label{eq10}
& &\ddot{w}_{ 1}= 
\frac{1}{w_{ 1}^3} +\frac{2N_1a_1}{\sqrt{2\pi }w_1^4}  +
\frac{2N_1^2K_3}{9\sqrt 3 \pi^3 w_1^7}+
\frac{4a_{12}m_1 N_2 w_1}{\sqrt \pi m_R (w_1^2+w_2^2)^{5/2}}\nonumber \\
& &+  \frac{24\sqrt 2}{25\sqrt 5} \frac{\alpha a_1^{5/2} N_1^{3/2}}{\pi^{5/4}w_1^{11/2}}
, \\ \label{eq13}
& &
\ddot{w}_{ 2}=
\frac{m_{1}}{4m_2w_{ 2}^3} +\frac{6\sqrt 3 m_1 (3 \pi^2 N_2)^{2/3 }}{25 \sqrt 5 m_2 \pi w_2^3}+
\frac{4a_{12}m_1 N_1 w_{ 2}  }{\sqrt \pi  m_R (w_1^2+w_2^2)^{5/2}}.
\end{eqnarray} 
The solution of the time-dependent 
equations (\ref{eq10}) $-$ (\ref{eq13}) gives the dynamics of the variational approximation. For static 
properties of the boson-fermion quantum ball, the time derivatives in these equations should be set equal to zero.

\begin{figure}[!t]
\begin{center}
\includegraphics[width=\linewidth]{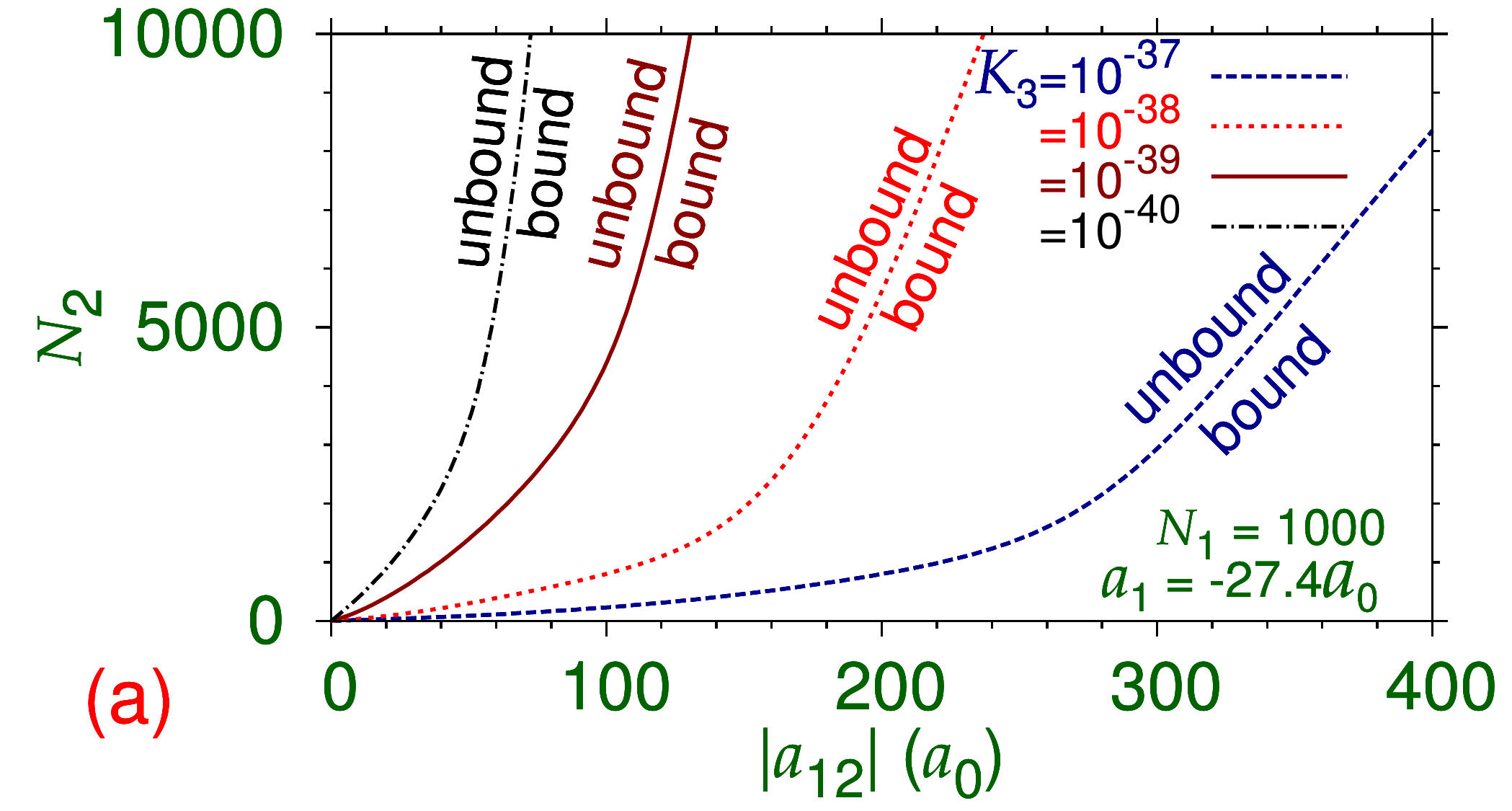}
\includegraphics[width=\linewidth]{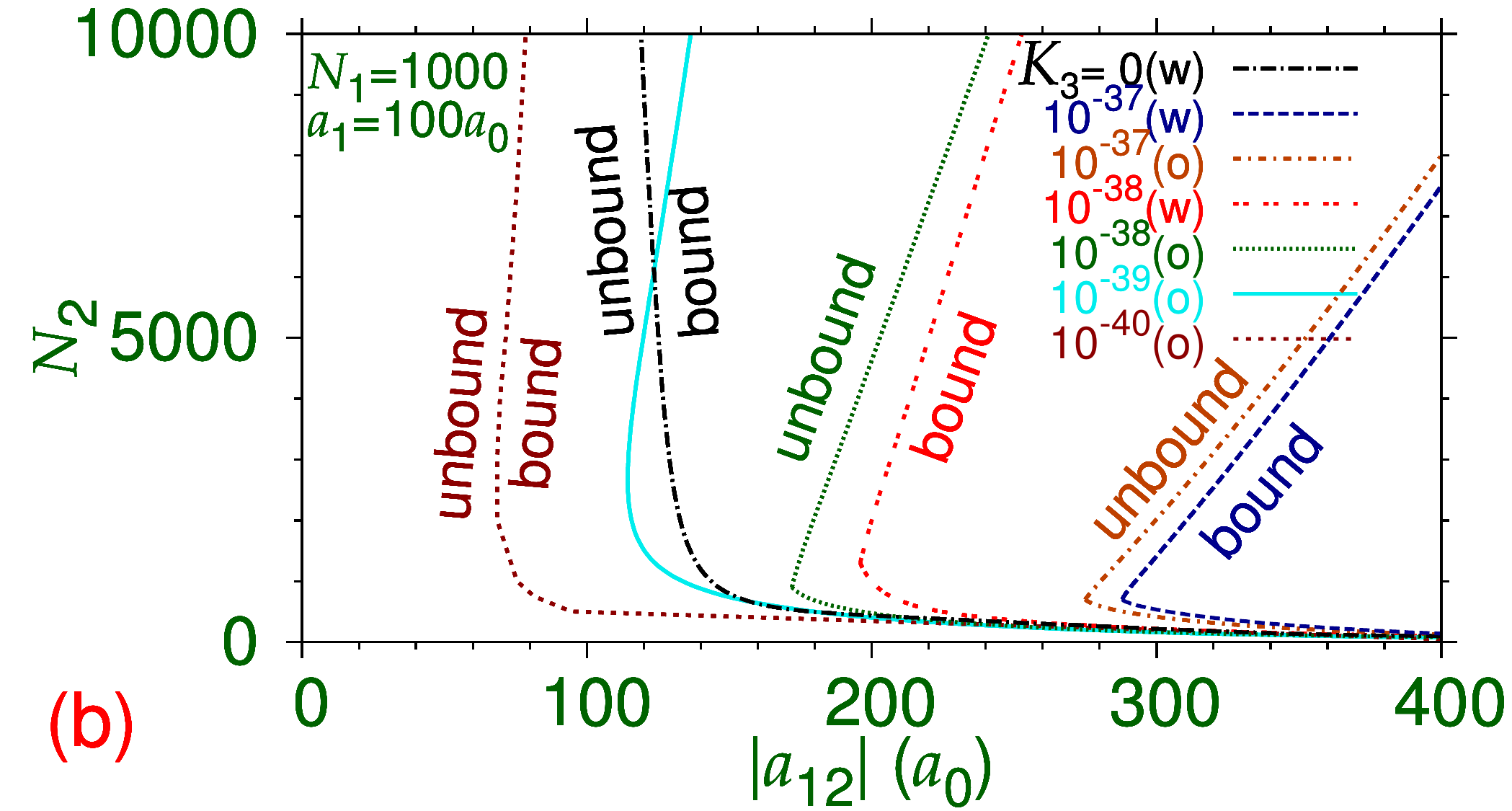}
 
\caption{ (Color online) Variational  $N_2-a_{12}$
 stability plot for the formation of boson-fermion $^7$Li-$^6$Li quantum ball of $N_1=1000$ bosons  
for { $K_3=0, 10^{-37}$ m$^6$/s, $10^{-38}$ m$^6$/s,  $10^{-39}$ m$^6$/s,   $10^{-40}$ m$^6$/s,} and for boson-boson scattering length (a) $a_1=-27.4a_0$ and 
(b) $100a_0$. In (b)   results are shown with (w) and without (o) the LHY correction term. The formation of the boson-fermion quantum ball is possible in the 
region to the right  of each line  marked ``bound".  No bound quantum ball is possible on the left side of the lines marked ``unbound".    
  }\label{fig1}
\end{center}
\end{figure}

The energy of the system is given by 
\begin{eqnarray}\label{energy}
 E
&= &
 \frac{3N_1}{4  w_{ 1}^2 } 
  +\frac{3N_2m_{1}}{16m_2   w_2^2} 
+\frac{N_1^2a_1}
{\sqrt{2\pi}w_{ 1}^3} +\frac{N_1^3K_3}{18\sqrt{3}\pi^3 w_1^6}
\nonumber\\ 
& + &\frac{8\sqrt 2}{25\sqrt 5} \frac{\alpha a_1^{5/2} N_1^{5/2}}{\pi^{5/4}w_1^{9/2}}
+\frac{9\sqrt 3 m_{1}(3 \pi^2 N_2)^{2/3} N_2}{50\sqrt 5  m_2\pi w_2^2} 
\nonumber\\
 & 
+&\frac{2a_{12}m_1N_1N_2}{\sqrt \pi m_R
 (w_1^2+w_2^2)^{3/2}}    .      
\end{eqnarray}  
The widths of the stationary state can be 
obtained from the solution of equations (\ref{eq10}) $-$ (\ref{eq13}) setting the time derivatives of the widths 
equal to zero.  This procedure is equivalent to a minimization of the energy (\ref{energy}), provided the 
stationary state   corresponds to a energy minimum.

\begin{figure}[!t]
\begin{center}\includegraphics[width=\linewidth]{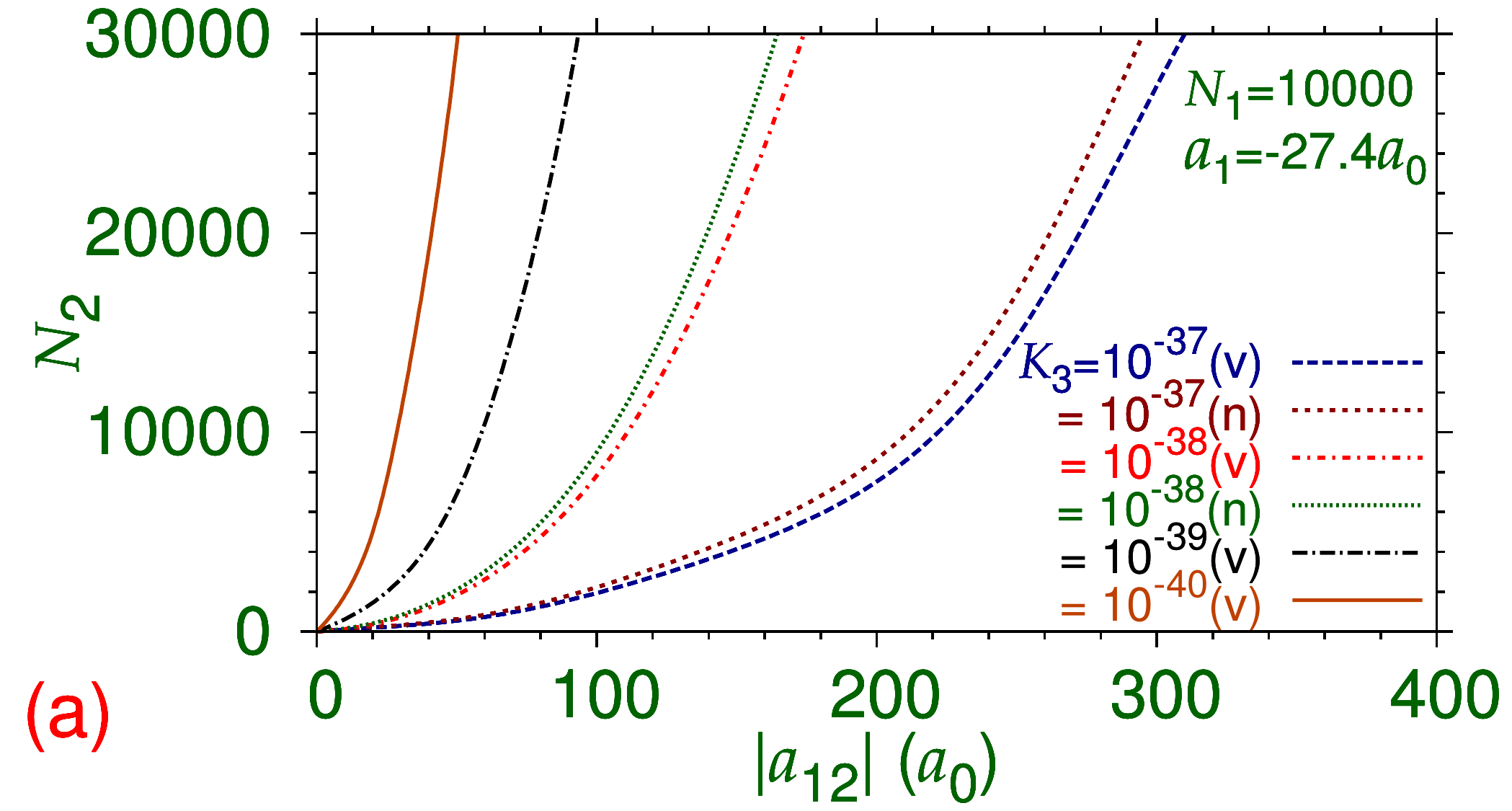}
\includegraphics[width=\linewidth]{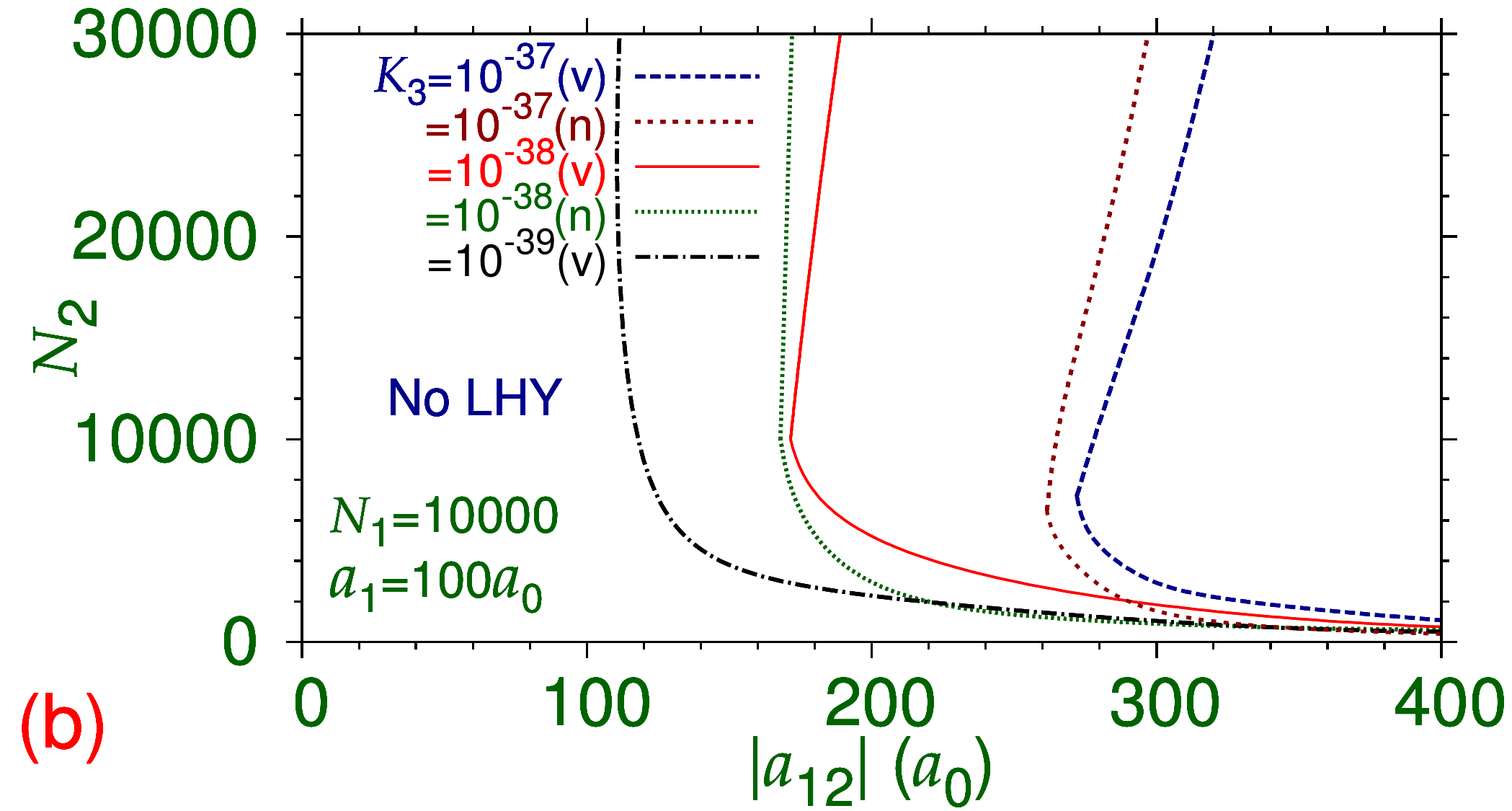}
\includegraphics[width=\linewidth]{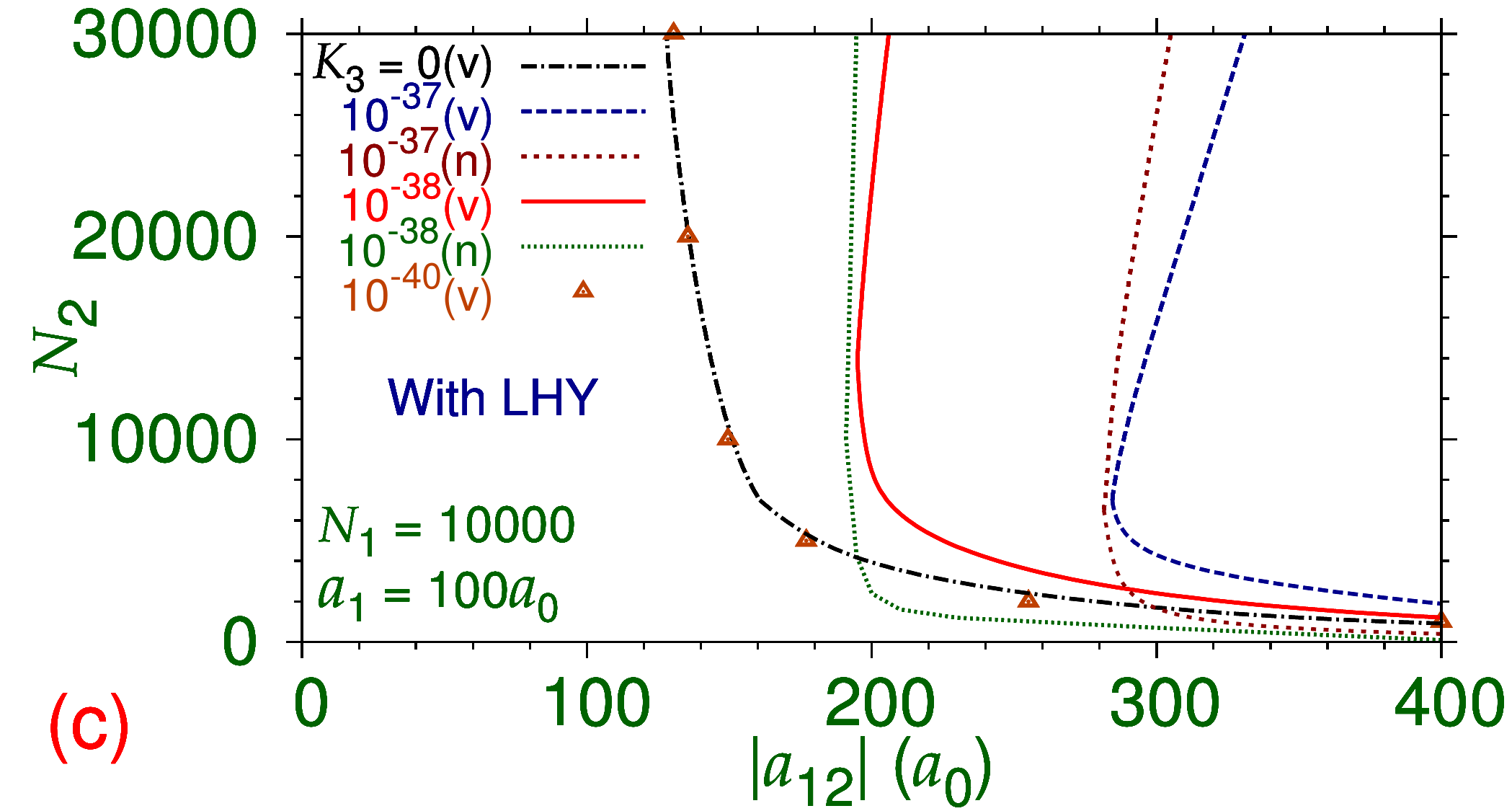}
  
\caption{ (Color online) Variational (v) and numerical (n)   $N_2-a_{12}$
 stability plot for the formation of boson-fermion $^7$Li-$^6$Li quantum ball of $N_1=10000$ bosons  
for {  different $K_3$ from 0 to $10^{-37}$ m$^6$/s} and for boson-boson scattering length (a) $a_1=-27.4a_0$,
(b)  $100a_0$  (without LHY correction), and  (c)  $100a_0$  (with LHY correction).   The formation of 
boson-fermion quantum ball is possible only in the right side of these lines. 
  }\label{fig2}
\end{center}
\end{figure}

\section{Numerical Results}

The
3D binary mean-field  equations (\ref{eq3}) and (\ref{eq4})   do
not have analytic solution and
different  numerical  methods,  such  as  split-step  Crank-
Nicolson \cite{CN} and Fourier spectral \cite{FS} methods, are used
for its solution. We solve these equations  numerically by the split-step Crank-Nicolson 
method using both
real- and imaginary-time propagation. 
  Imaginary-time  simulation  is
employed to get the lowest-energy bound state of the boson-fermion quantum ball, while the real-time simulation is to be used to study the dynamics using the initial profile obtained in
the imaginary-time propagation \cite{24}.  There are different
C and FORTRAN programs for solving the GP equation
\cite{CN,24} and one should use the appropriate one.  In the
imaginary-time propagation the initial state was taken as
in   \ref{anz}  and the width
$w_i$
set equal to the variational
widths.  The convergence
will be quick if the guess for the widths
$w_i$
is close to the
final converged width.

We consider the boson-fermion $^7$Li-$^6$Li mixture in this study with the experimental 
scattering length $a_1=a(^7$Li$)=-27.4a_0$. This negative scattering length imply intra-species 
attraction in  $^7$Li.  We also consider $a_1=100a_0$: it is also possible to have a boson-fermion 
quantum ball for for a repulsive boson-boson interaction and an attractive boson-fermion 
interaction. In the latter case the LHY correction  is also effective.  
The fermions are considered to be in the weak-coupling BCS limit 
without any inter-species interaction between spin-up and -down fermions.  
The yet unknown inter-species scattering length $a_{12}$ is taken as a variable. The variation of  
$a_{12}$  and $a_1$ can be achieved experimentally by the optical \cite{fesho}  and magnetic  \cite{fesh}
Feshbach resonance techniques.
We consider  
 the length scale $l_0=1$ $\mu$m and consequently, the  time scale $t_0=
0.11$ ms.

We find that a  boson-fermion   $^{7}$Li-$^6$Li  quantum ball 
is achievable for a moderately attractive inter-species attraction (negative $a_{12}$) and  for 
appropriate values of the number of atoms, for both attractive and repulsive boson-boson interaction. 
We illustrate in figure \ref{fig1}  
the $N_2-|a_{12}|$  variational  stability plots for a boson-fermion quantum ball for boson-boson scattering lengths   (a)
$a=-27.4a_0$ and (b) $a=100a_0$, for $N_1=1000$ and {  $K_3=0, 10^{-37}$  m$^6$/s,  $10^{-38}$ m$^6$/s,  $10^{-39}$ m$^6$/s, and  $10^{-40}$ m$^6$/s.  We find that a boson-fermion quantum ball can be formed for different non-zero values of $K_3$ with other parameters unchanged. However, a reduced $K_3$ value implies an increased net attraction, thus resulting in a more tightly bound boson-fermion quantum ball of reduced size.}
In the case of 
repulsive boson-boson interaction we also included the LHY correction. The stability plots are qualitatively 
different for attractive and repulsive boson-boson interaction. For an attractive boson-boson interaction a boson-fermion 
quantum ball can be formed for a weakly attractive boson-fermion interaction. However, for a repulsive boson-boson interaction 
a boson-fermion quantum ball can be formed for the boson-fermion attraction above a critical value.

\begin{figure}[!t]

\begin{center}
\includegraphics[width=.48\linewidth]{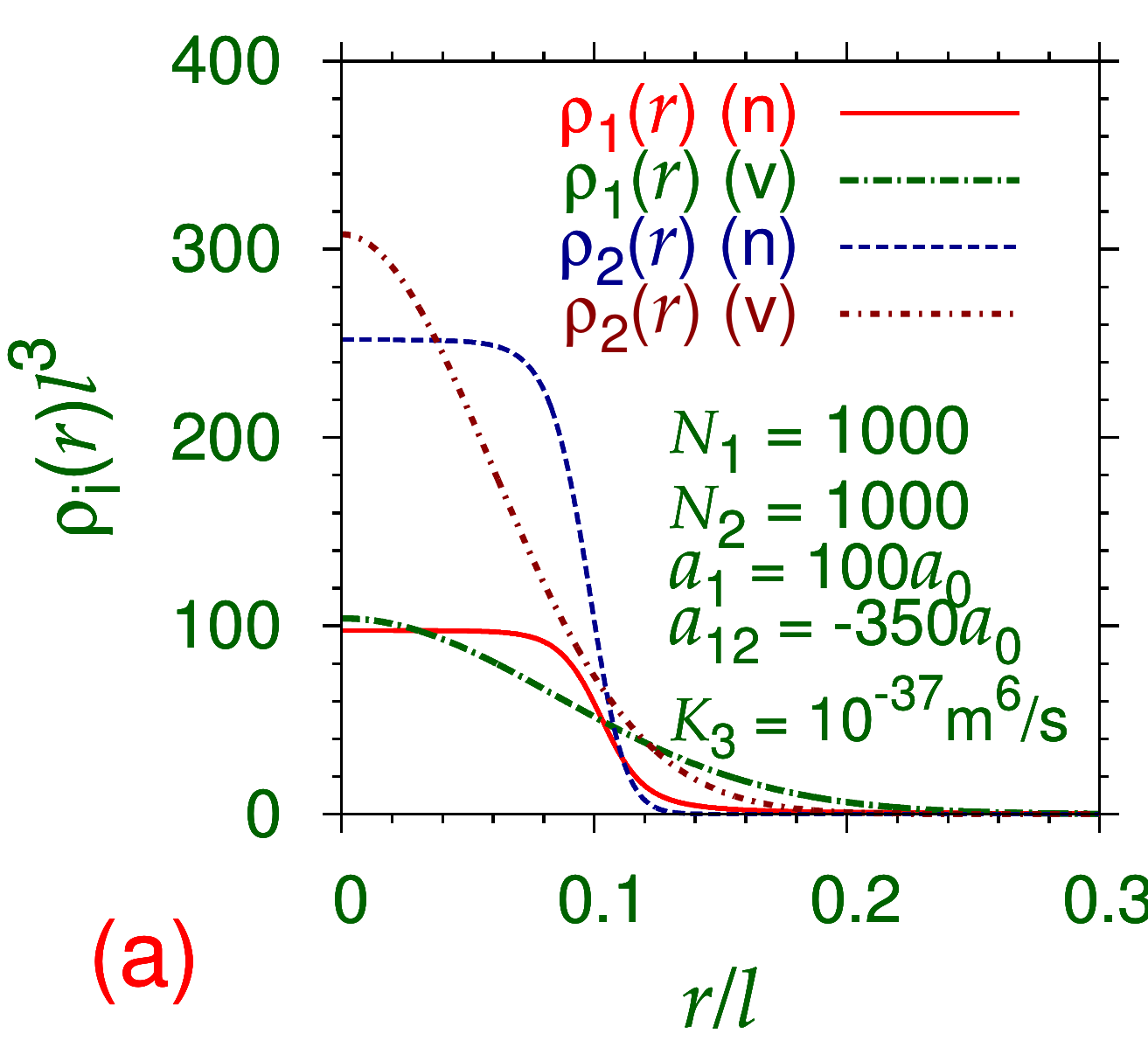}
\includegraphics[width=.48\linewidth]{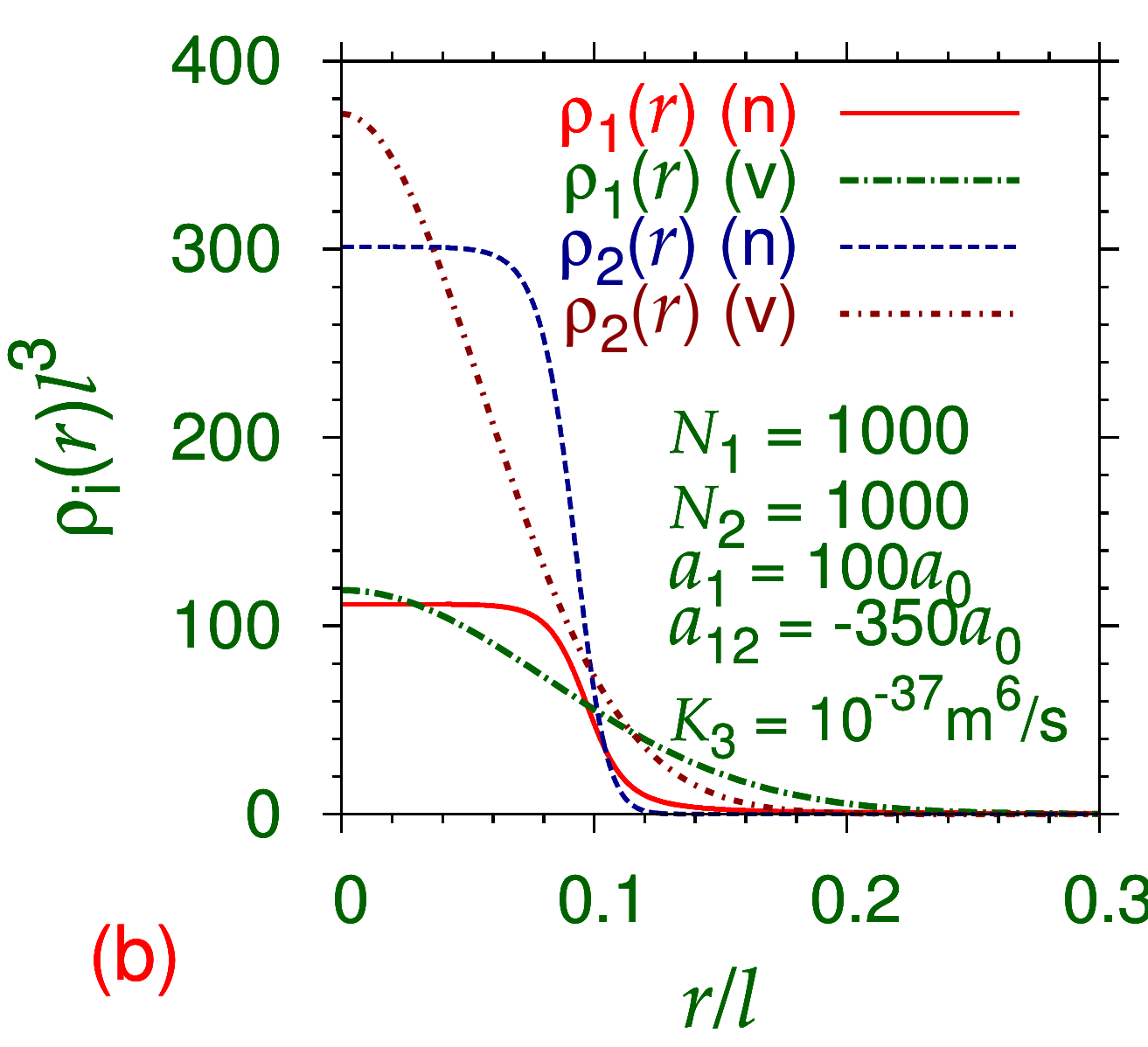}
\includegraphics[width=.48\linewidth]{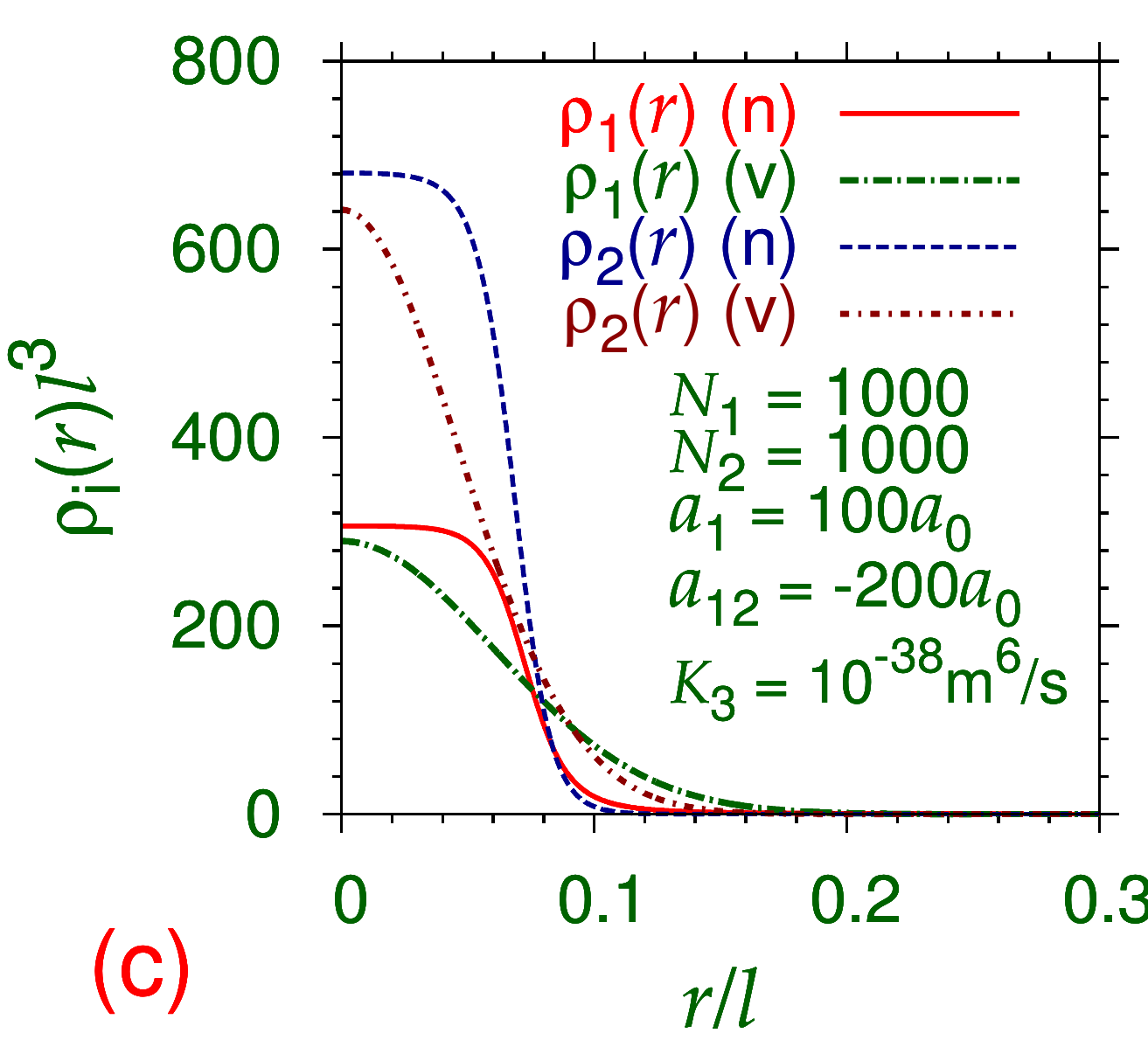}
\includegraphics[width=.48\linewidth]{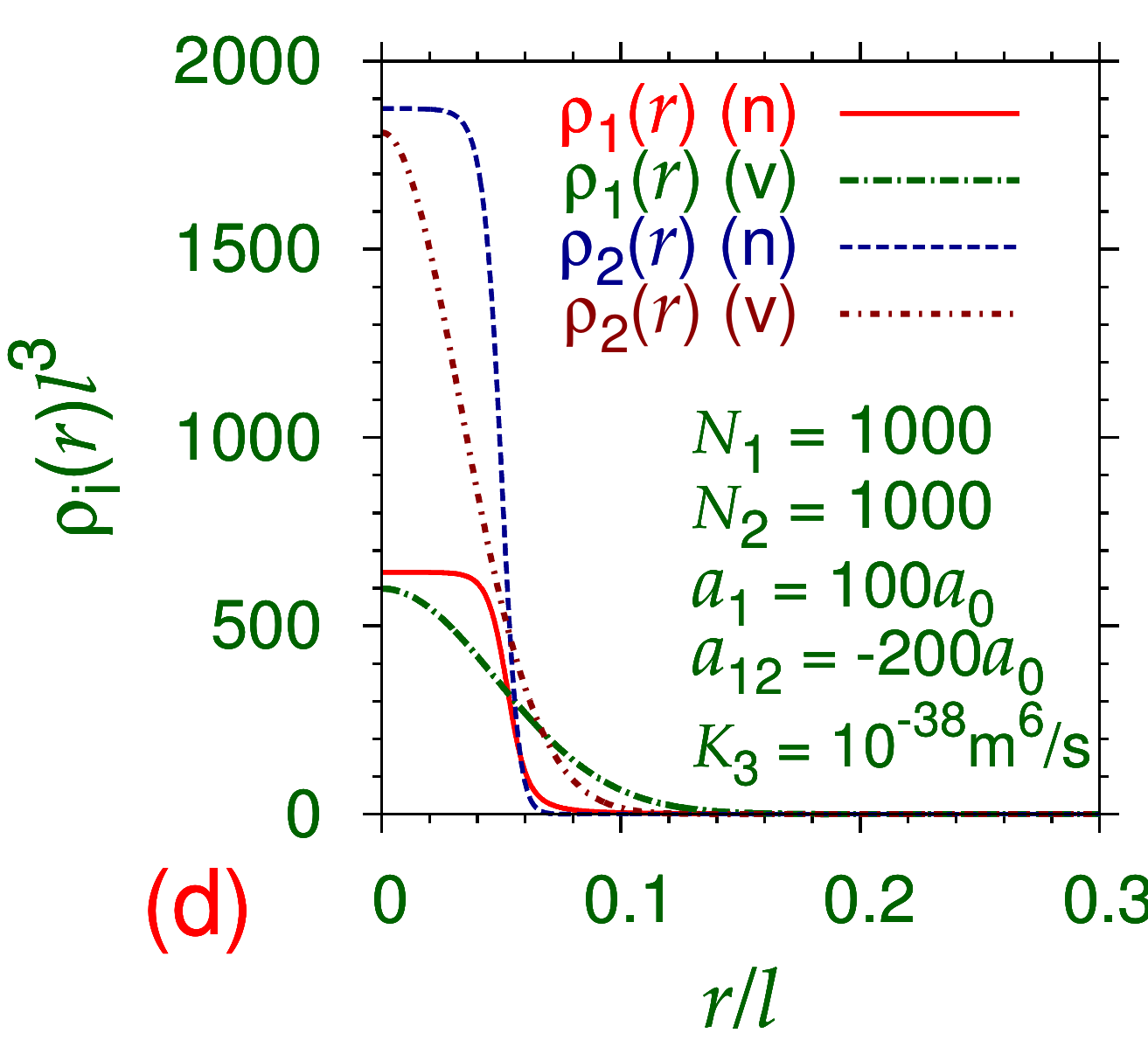}
\includegraphics[width=.48\linewidth]{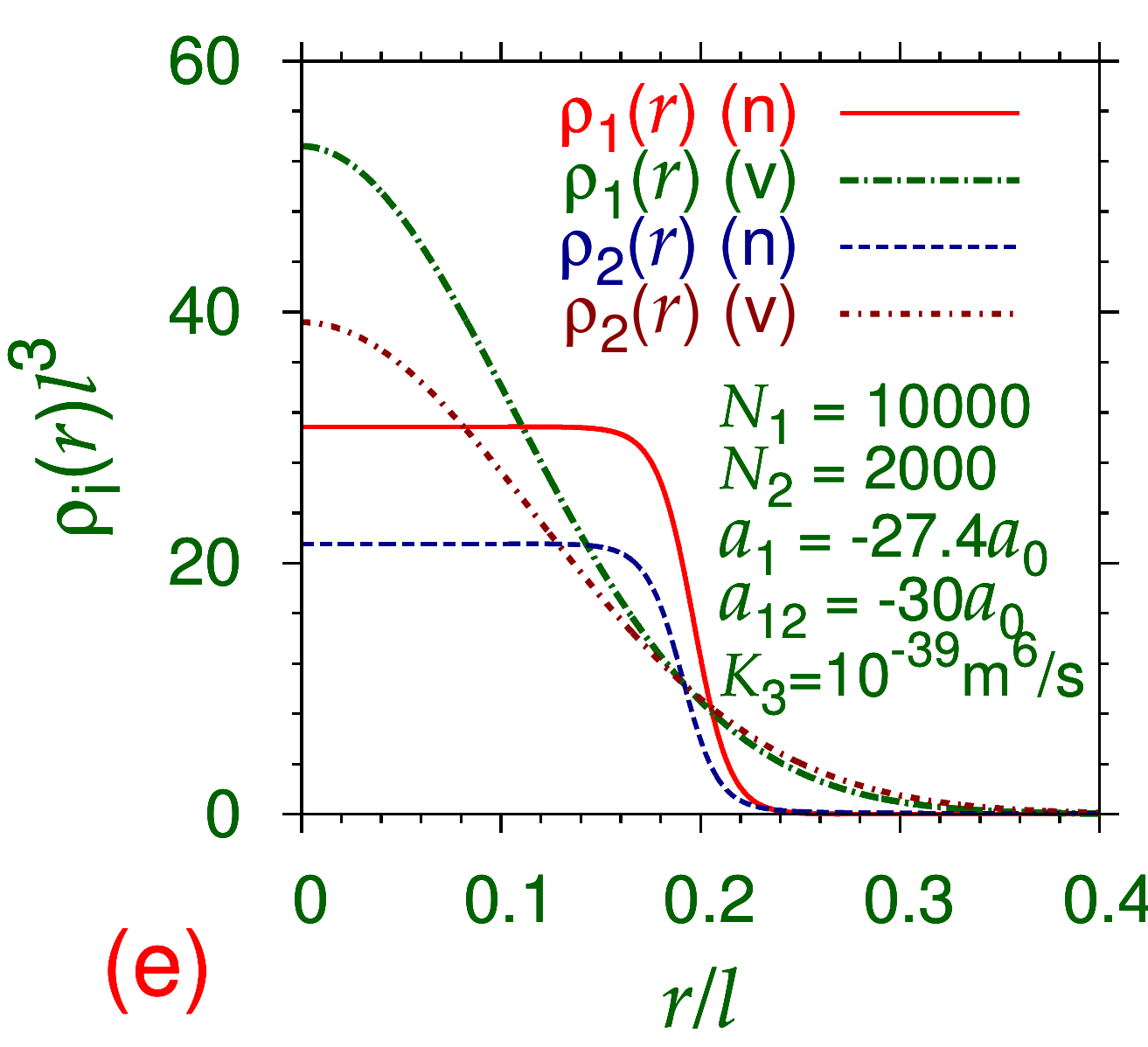}
\includegraphics[width=.48\linewidth]{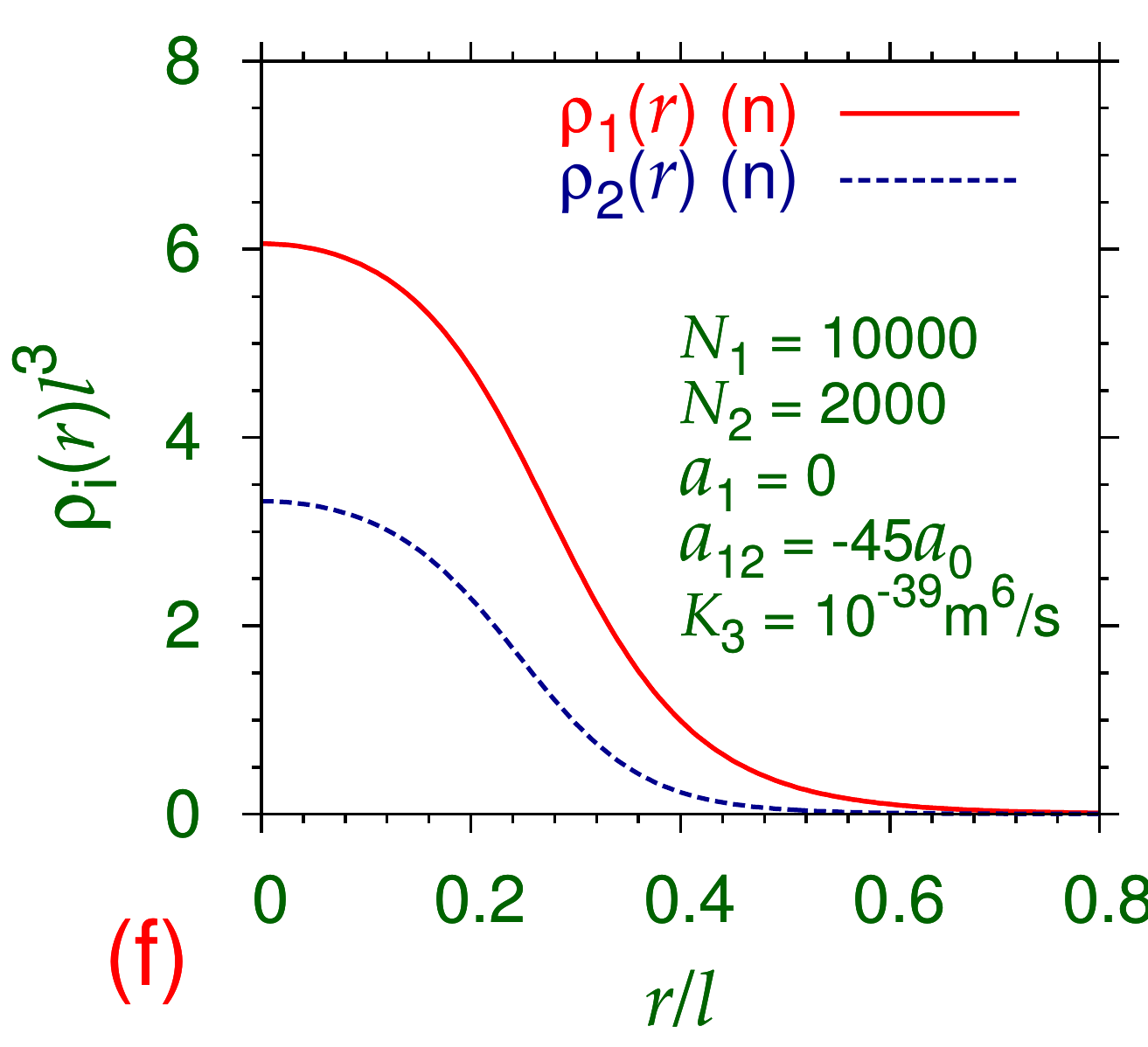}
\caption{ (Color online)Variational (v) and numerical (n) densities $\rho_i=|\psi_i|^2$ 
of the bosons and fermions for different sets of parameters and for (a)   $N_1=N_2=1000, a_1=100a_0, a_{12}=-350a_0, K_3=10^{-37}$ m$^ 6$/s with LHY correction, (b)  $N_1=N_2=1000, a_1=100a_0, a_{12}=-350a_0, K_3=10^{-37}$ m$^ 6$/s without LHY correction,{  (c) $N_1= N_2=1000, a_1=100a_0, a_{12}=-200a_0, K_3=10^{-38}$ 
 m$^ 6$/s with LHY correction,   (d) $N_1= N_2=1000,  a_1=100a_0, a_{12}=-200a_0,
 K_3=10^{-38}$ m$^ 6$/s without  LHY correction, (e)  $N_1=10000, N_2=2000, a_1=-27.4a_0, a_{12}=-30a_0,  K_3=10^{-39}$  m$^ 6$/s, (f) $N_1=10000, N_2=2000,a_1=0,  a_{12}=-45a_0,  
K_3=10^{-39}$ m$^ 6$/s.}  The plotted quantities in this and following figures
 are dimensionless. The unit of length $l$ in all figures is $l=1$ $\mu$m. 
    }\label{fig3}
\end{center}

\end{figure}

\begin{figure}[!t]
\begin{center}
\includegraphics[width=\linewidth]{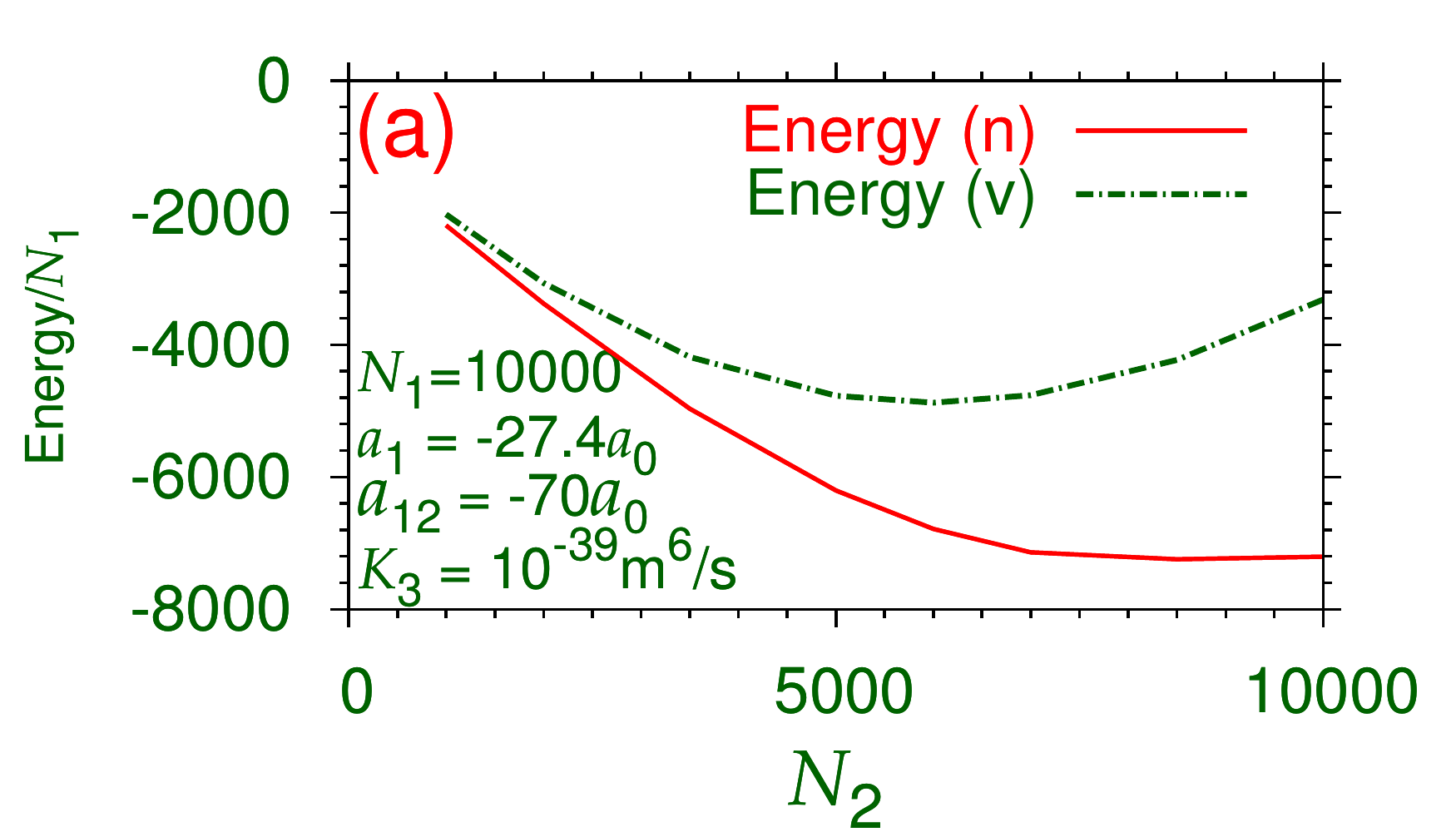}
\includegraphics[width=\linewidth]{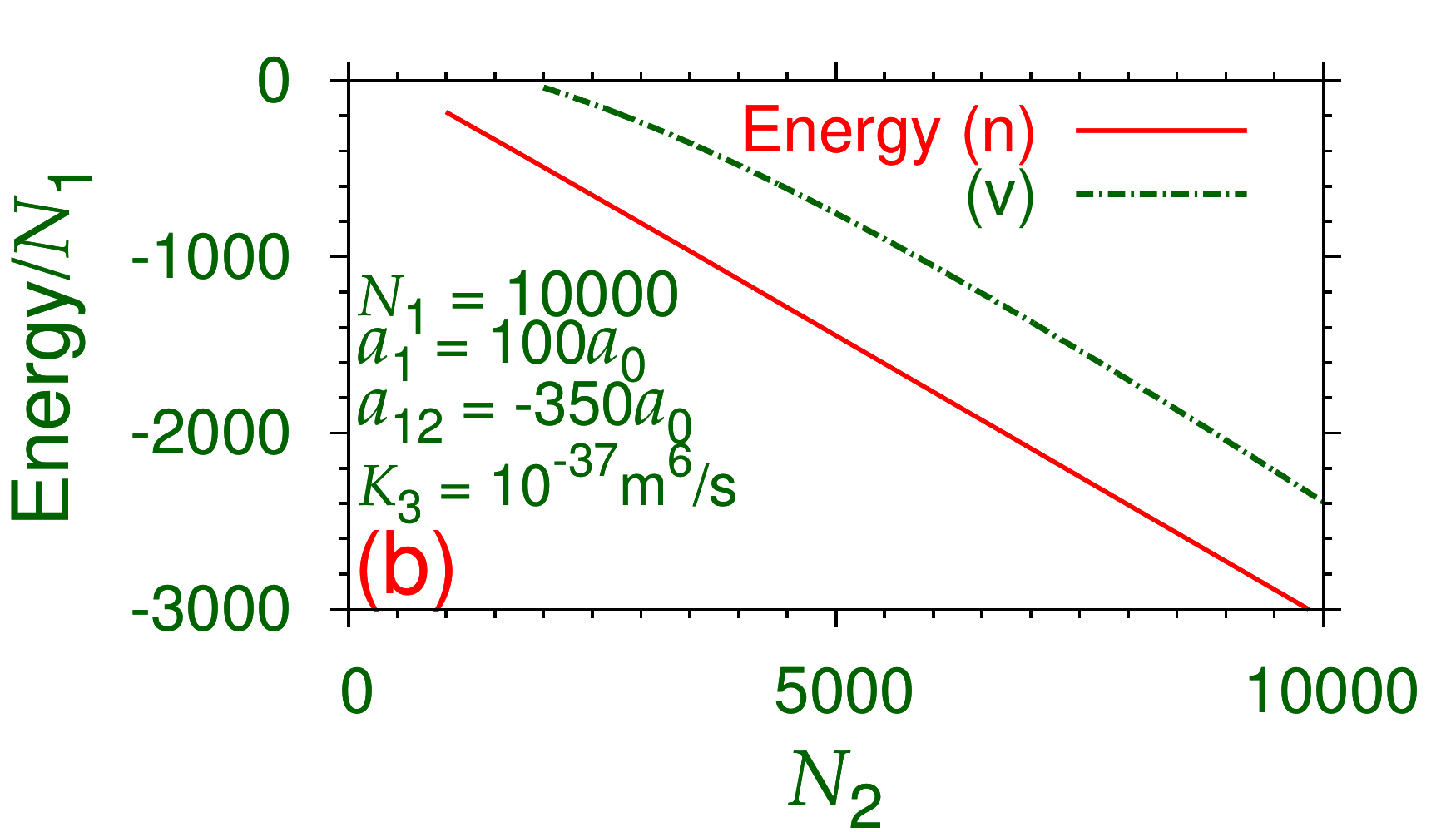} 
\caption{ (Color online) Variational (v) and numerical (n) energies versus $N_2$ for $N_1=10000,$ { 
and  (a) $a_1=-27.4a_0, a_{12}= -70a_0, K_3=10^{-39}$ m$^6$/s  and (b)  $a_1=100a_0, a_{12}= -350a_0,
 K_3=10^{-37}$ m$^6$/s  }
without LHY correction. The unit of energy is $\hbar^2/(m_1l^2)$.
}\label{fig4}
\end{center}

\end{figure}

\begin{figure}[!b]
\begin{center}
\includegraphics[width=\linewidth]{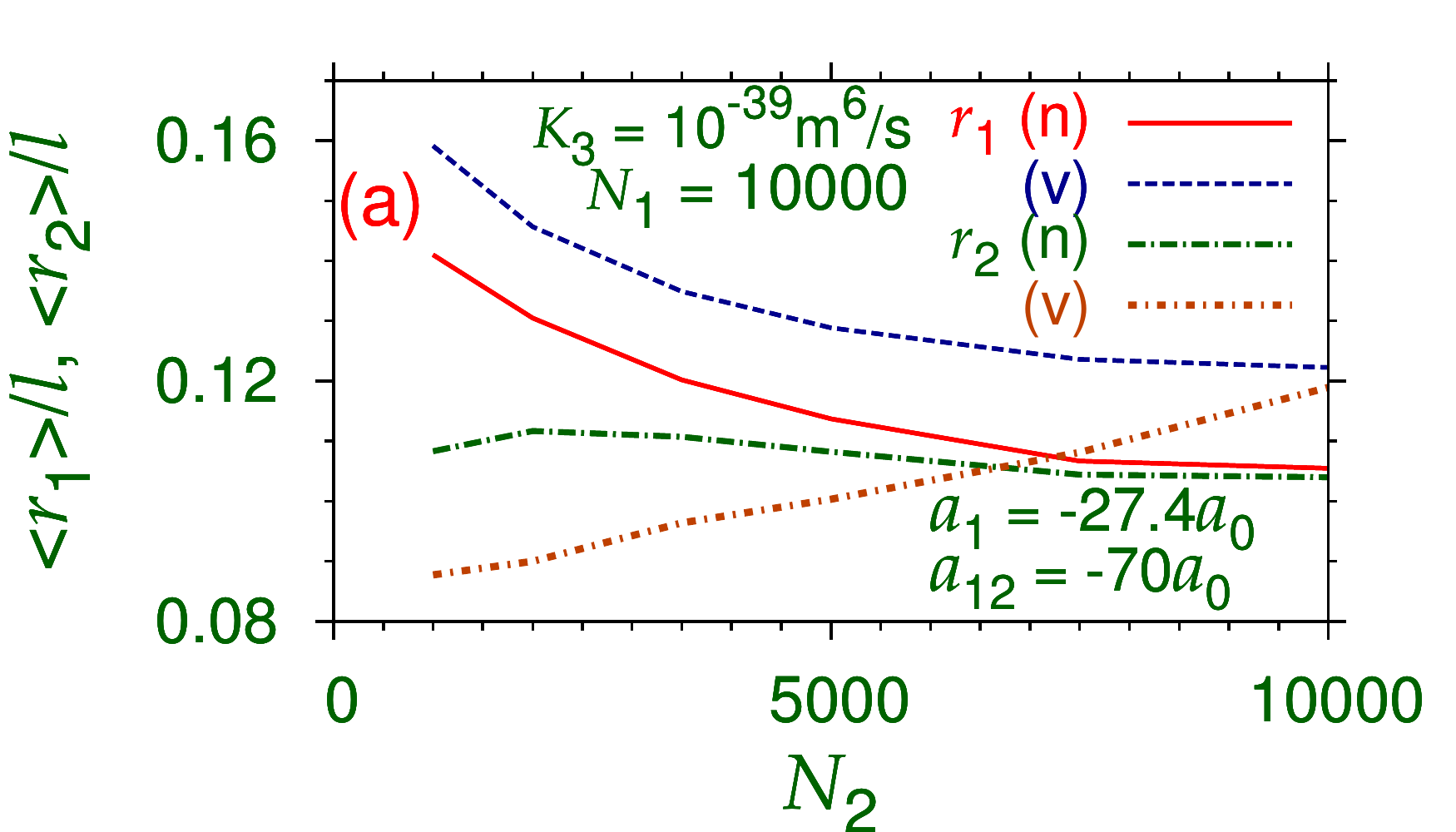}
\includegraphics[width=\linewidth]{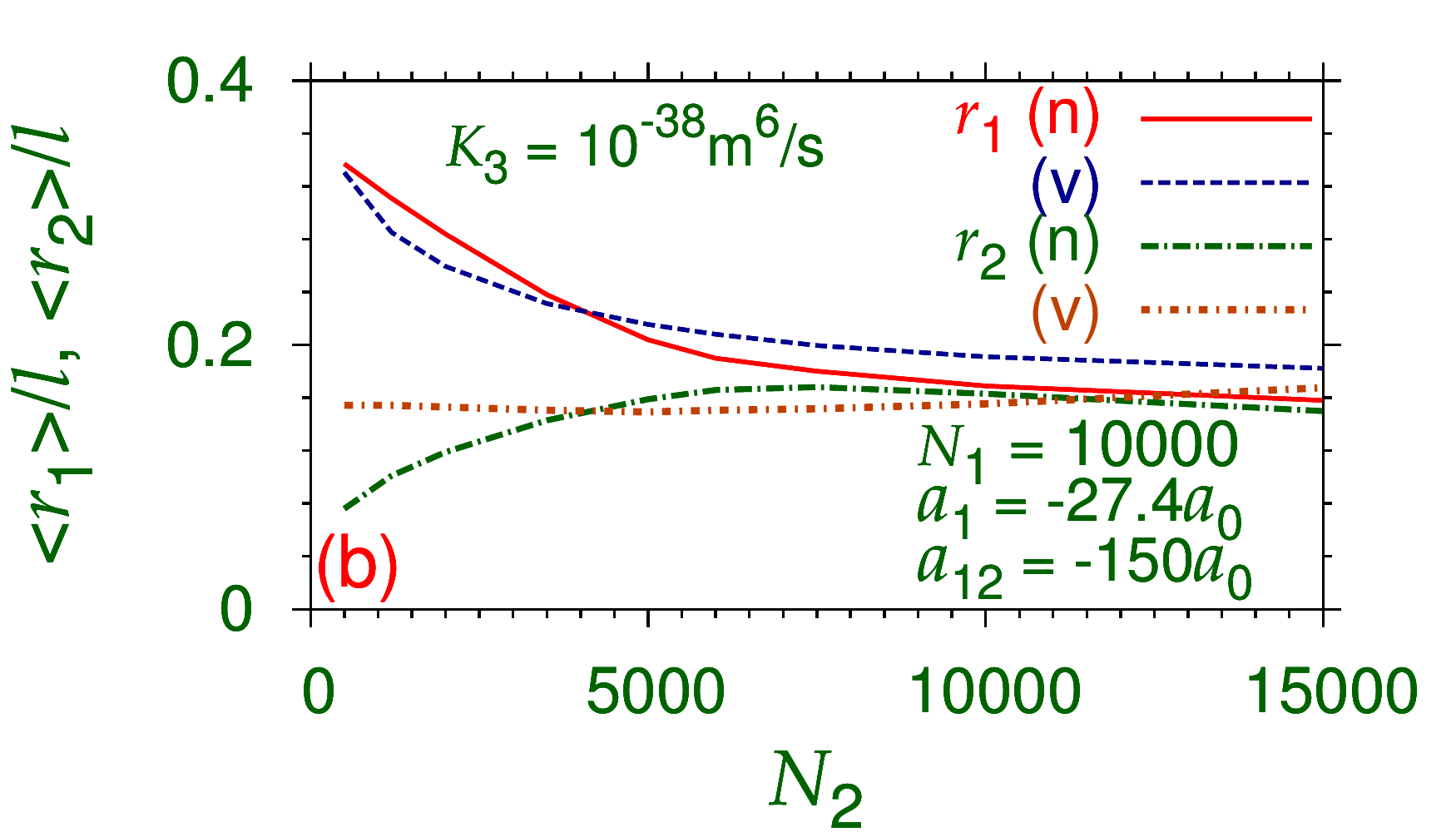} 
\caption{ (Color online) Variational (v) and numerical (n) rms sizes $\langle r_1 \rangle$ and  $\langle r_2 \rangle$ versus $N_2$ for $N_1=10000,a_1=-27.4a_0$ for (a)
{  $ K_3=10^{-39}$ m$^6$/s , $a_{12}= -70a_0$} and (b)   $ K_3=10^{-38}$ m$^6$/s , $a_{12}= -150a_0$.
   }\label{fig5}
\end{center}

\end{figure}

In figure \ref{fig2} we display  similar  variational and numerical  $N_2-|a_{12}|$ stability plots for $N_1=10000$
for (a) $a_1=-27.4a_0$, (b) $a_1=100a_0$ (without LHY correction), and (c) $a_1=100a_0$ (with LHY correction) { for different $K_3$ values}.   The numerical results for the stationary quantum balls 
in figures \ref{fig2}-\ref{fig5} are obtained by imaginary-time simulation.
The formation of the boson-fermion quantum ball is possible on the right of the plotted lines in figures \ref{fig1} and \ref{fig2}.  There is not enough 
attraction on the left side of these lines to bind such a quantum ball.
The numerical lines lie on the left of the variational lines showing a larger domain for the formation of the quantum balls. 
This is a consequence of the fact that the variational energies set an upper bound on the actual energy.   
Also the stability lines with the LHY correction correspond to  an increased repulsion and the stability lines move towards right, 
viz. 
figures \ref{fig2}(b) and (c) implying a reduced domain in the parameter space for the formation of boson-fermion quantum ball.

We used a Gaussian  ansatz   for the variational approximation, which is the eigenfunction of a harmonic oscillator. 
This ansatz should work well in the presence of a harmonic trap with small values of nonlinear interaction. 
In the present case,  there is no harmonic trap and the nonlinearities could be quite large. Hence the variational 
approximation is not expected to be good in general.  We have seen that the variational approximation has yielded 
qualitatively correct result for the stability plots, viz. figures \ref{fig1}  and \ref{fig2}. 
To test how well the variational approximation can yield the density  profiles, we have compared in figure \ref{fig3}
the variational and numerical densities of the boson-fermion quantum ball for different cases for (a)-(b) $K_3=10^{-37}$  m$^6$/s, 
{  (c)-(d) $10^{-38}$  m$^6$/s, and (e)-(f) $10^{-39}$  m$^6$/s}.
For repulsive boson-boson interaction,  we have also included the LHY correction term in figures \ref{fig3}(a) and (c).     The inclusion of LHY 
correction implies more repulsion: consequently, the density profiles are more extended in space with  smaller central 
densities in these plots. 
In all cases the numerical densities are very different from a Gaussian shape. Considering that there is no harmonic trap in the model, the agreement between the variational and numerical results is quite satisfactory.

Now we compare the variational and numerical  energies of the boson-fermion quantum ball  versus number of fermions in figure \ref{fig4} 
for $N_1=10000$  and for { (a) $a_1=-27.4a_0, a_{12}=-70a_0$,  $K=10^{-39}$ m$^6$/s} and (b)    $a_1=100a_0, a_{12}=-350a_0,$  $K=10^{-37}$ m$^6$/s. {  The variational energies are are always larger than the numerical energies.}
In figure \ref{fig5} we plot the root-mean-square (rms) sizes
$\langle r_1 \rangle$ and  $\langle r_2 \rangle$
of bosons and fermions versus $N_2$ for $N_1=10000, a_1=-27.4a_0 $
and for {  (a) $a_{12} = -70a_0,  K=10^{-39}$ m$^6$/s } and (b) 
 $a_{12} = -150a_0,  K=10^{-38}$ m$^6$/s. The agreement between the variational and numerical  results is reasonable in both cases.

\begin{figure}[!t]
\begin{center}
\includegraphics[width=\linewidth]{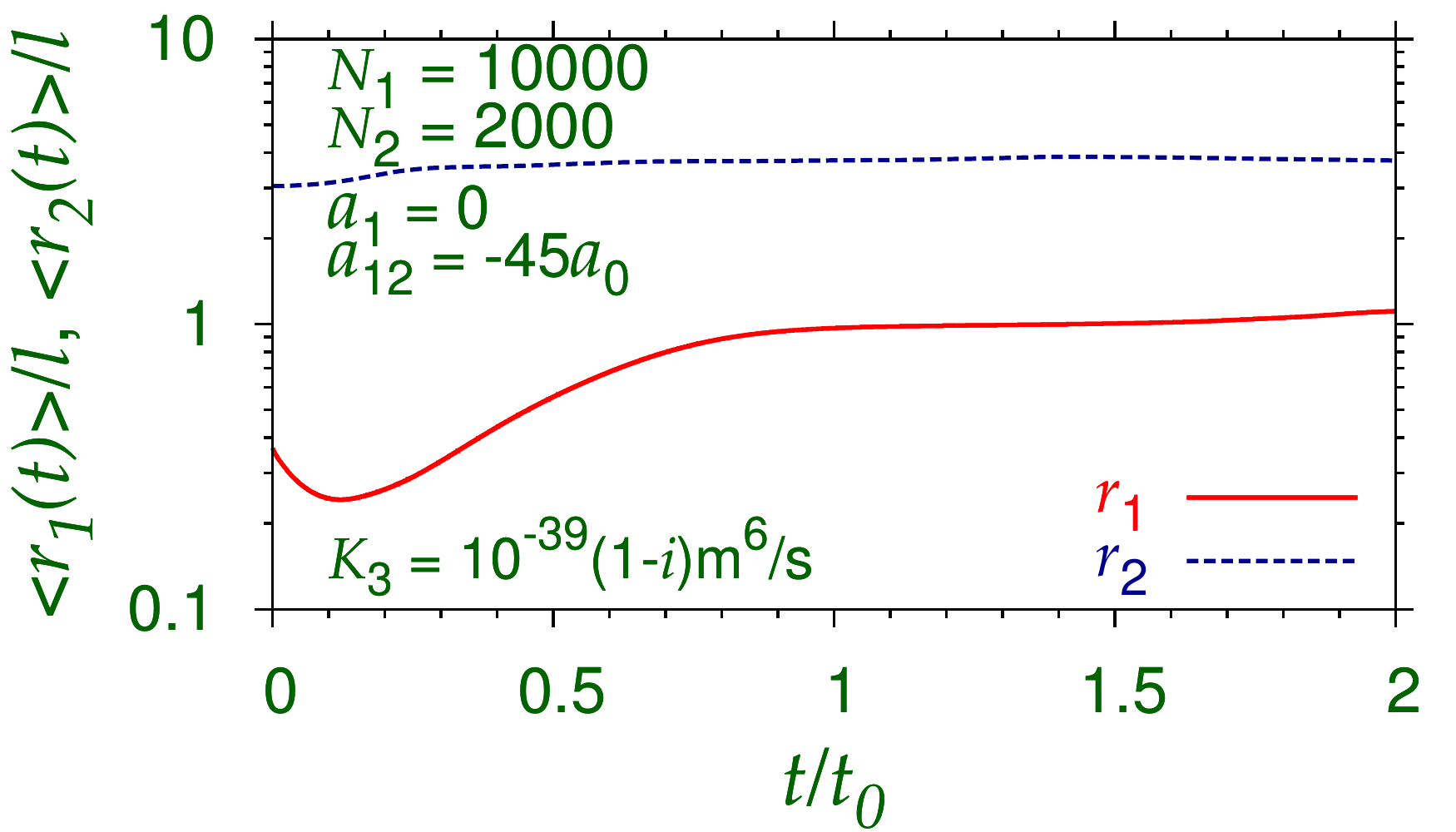} 
\caption{ (Color online)  Dynamical oscillation of the rms sizes $\langle r_1, r_2\rangle$ upon real-time propagation 
of the boson-fermion $^7$Li-$^6$Li 
quantum ball of figure \ref{fig3}(f) 
 prepared by imaginary-time propagation in a harmonic trap of frequency $\omega =2 \pi \times 1443$ Hz. The plotted quantities are dimensionless. The harmonic oscillator length $l=1$ $\mu$m, and the time scale $t_0= 0.11$ ms. 
   }\label{fig6}
\end{center}

\end{figure}

   We have seen that these boson-fermion quantum balls are very tightly bound, viz. the large energy/boson in figure \ref{fig4}. The best way to observe these solitons is to prepare  these boson-fermion quantum balls in a harmonic trap and then remove the trap. To this end we numerically prepared by imaginary-time propagation 
a boson-fermion quantum ball {  for $N_1=10000, N_2=2000, a_1= 0, a_{12}=-45a_0, K_3=10^{-39}$ m$^6$/s}
in a harmonic trap of frequency 
$\omega =2\pi \times 1443 $ Hz which corresponds to a harmonic oscillator length $l\equiv \sqrt{\hbar/m_1\omega}= 1$ $\mu$m 
for $^7$Li  atoms. 
Then we performed real-time propagation without   a trap with the same parameters using the imaginary-time state as the initial state. {  In this simulation we have included an imaginary part to the three-body term $K_3$ to take 
into account 
the three-boson loss. There is  estimate of three-body loss for $^7$Li atoms \cite{K3} for different values of scattering length $a_1$, although its  value for $a_1=0$ is not given there. We take the three-body loss 
$K_3= -i10^{-39}$ m$^6$/s, which is the average value away from the nearby Feshbach resonance where $a_1\to\pm  \infty.$  In the present real-time simulation we use $K_3= (1-i)10^{-39}$ m$^6$/s, which takes into account a realistic three-body loss.  Due to the presence of the absorptive term in $K_3$,  the number of bosons decay with time. Nevertheless, a smaller number of bosons is enough to keep the fermions bound  
due to the attractive boson-fermion interaction. In figure \ref{fig6} we plot the rms sizes of the bosons and fermions versus time.  A practically constant rms size of the fermions guarantee the stability of the quantum ball.  Due to a sudden introduction of the three-body loss term at $t=0$ some  disturbance is created in the quantum ball, as the initial state obtained by imaginary-time simulation is not an eigenstate of the absorptive Hamiltonian with three-body loss.  
The large values of the rms radius $r_2$ of fermions result due to some small noise at large values of $r$, although the quantum ball  remain localized near the center.   }

\section{Summary and Discussion} 
  
We demonstrated the possibility of the creation of a stable, stationary, self-bound  super-fluid 
boson-fermion   quantum ball  under attractive inter-species interaction using a variational and a
numerical solution of a mean-field model. The boson-boson interaction could be attractive or repulsive. 
The collapse is avoided by a three-boson interaction and/or  a LHY correction to the 
two-boson interaction.    
The static properties of the boson-fermion quantum
ball are studied by the variational approximation and a
numerical  imaginary-time  solution  of  the  mean-field model.
The  dynamics is   studied  by  a  real-
time solution of the same using the imaginary-time solution as input.
The numerical and variational results for the rms radii, densities, and energies 
of the boson-fermion   quantum ball are in agreement with each other.

The binary quantum  ball is very tightly bound even for  a  small  three-boson interaction and/or  a  small LHY correction, hence should be easy to observe in a laboratory like the boson-boson quantum ball \cite{leticia}.  We demonstrate a possible practical mean for its formation.  A boson-fermion mixture should be kept in a harmonic trap of harmonic oscillator length of few microns  with parameters appropriate for the formation of a quantum ball.  Actually, one of the easiest way of achieving a degenerate fermion gas is by sympathetic cooling in  a boson-fermion mixture, such as in $^7$Li-$^6$Li \cite{sc}.  Such a mixture should be used to create the boson-fermion quantum ball.  
  Usually the size of the quantum ball will be 
much smaller than the  harmonic oscillator length, indicating that the harmonic trap has no effect on the formation of the
quantum ball. Consequently, the removal of the harmonic trap will have marginal effect on the quantum ball. To demonstrate this in numerical simulation, we form a quantum ball by imaginary-time propagation in a harmonic trap. Then we use the state so formed in a real-time propagation without a harmonic trap maintaining all other parameters the same.  Bounded  values of the rms radii in real-time propagation illustrates the stability of the quantum ball as well as the feasibility of its creation in a laboratory.

  \section*{Acknowledgments} 


The study was partially   supported by the Funda\c c\~ao de
Amparo \`a Pesquisa do Estado de S\~ao Paulo FAPESP (Brazil)
under Projects No. 2012/00451-0 and No. 2016/01343-7 and
also by the Conselho Nacional de Desenvolvimento Cient\'ifico
e Tecnol\'ogico (Brazil) under Project No. 303280/2014-0.

\end{document}